\documentclass[prb,twocolumn,showpacs,amsmath,amssymb]{revtex4}

\usepackage{graphicx}
\usepackage{dcolumn}
\usepackage{bm}
\usepackage{color}

\begin{document}

\newcommand{\bn}{{\bf n}}
\newcommand{\bp}{{\bf p}}   
\newcommand{\br}{{\bf r}}
\newcommand{\bk}{{\bf k}}
\newcommand{\bv}{{\bf v}}
\newcommand{\brho}{{\bm{\rho}}}
\newcommand{\wk}{\omega_{\bf k}}
\newcommand{\nk}{n_{\bf k}}
\newcommand{\eps}{\varepsilon}
\newcommand{\la}{\langle}
\newcommand{\ra}{\rangle}
\newcommand{\be}{\begin{eqnarray}}
\newcommand{\ee}{\end{eqnarray}}
\newcommand{\intl}{\int\limits_{-\infty}^{\infty}}
\newcommand{\dE}{\delta{\cal E}^{ext}}
\newcommand{\SE}{S_{\cal E}^{ext}}
\newcommand{\dsp}{\displaystyle\vphantom{\Bigl|}}
\newcommand{\Dsp}{\displaystyle\vphantom{\Biggl|}}
\newcommand{\phit}{\varphi_{\tau}}
\newcommand{\EF}{{\cal E}_F}
\newcommand{\fn}{f^{(0)}}
\newcommand{\Da}{D_\alpha}
\newcommand{\tm}{\tau_m}
\newcommand{\p}{\varphi}

\title{ Electron-electron scattering and magnetoresistance of ballistic microcontacts}

\author{ K. E. Nagaev }
\author{ T. V. Kostyuchenko }
\affiliation{Institute of Radioengineering and Electronics,  Mokhovaya 11-7, Moscow, 125009 Russia}

\date{\today}

\begin{abstract}
Using a semiclassical Boltzmann equation, we calculate corrections to the Sharvin conductance of a wide
2DEG ballistic contact that result from an electron--electron scattering in the leads. These corrections are 
dominated by collisions of electrons with nearly opposite momenta that come from different reservoirs. They
are positive, 
{increase with temperature,} and are strongly suppressed by a magnetic field. We argue that this suppression may be responsible for an anomalous positive magnetoresistance observed in a recent experiment.
\end{abstract}


\maketitle

\section{Introduction}

It is well known that narrow constrictions in 2D or 3D electron gases exhibit a finite electric resistance
even if their size is much smaller than the electron mean free path.\cite{Sharvin} In terms of the 
Landauer--B\"uttiker formalism, this resistance is due to a total backscattering of electrons in most of quantum channels of the electrodes.\cite{Imry} If a transverse magnetic field is applied to a constriction in a 2D gas, a suppression of geometrical backscattering  results in a temperature-independent negative magnetoresistance.\cite{Beenakker91}

In a recent paper, Renard et al. \cite{Renard08} studied electric transport in wide 2D quantum contacts formed of high-mobility GaAs heterostructures. In a zero magnetic field, the authors observed a positive correction to the conductance that increased linearly with temperature. In addition to this, they noticed an unusual behavior of magnetoresistance in low fields. The low-field magnetoresistance was positive and increased with temperature. In higher fields, it crossed over to a negative one, so that the $R(H)$ curves exhibited a temperature-dependent maximum. The authors attributed this temperature dependence to electron--electron interactions, but the specific 
mechanism of it was not clear. The goal of the present paper is to establish it.

There are several theories that address effects of electron--electron interaction in magnetoresistance of a uniform 2D electron gas.
In the semiclassical approximation, these interactions do not affect electric resistivity of macroscopically homogeneous conductors with a parabolic spectrum because of momentum conservation.
They are known to contribute to the resistivity only if the translational invariance of the conductor is broken by impurities and a quantum interference between the interactions and impurity scattering takes place.\cite{Altshuler85} There are several theories describing the magnetoresistance of a uniform 2D electron gas that results from these quantum effects. Gornyi and Mirlin \cite{Gornyi83,Gornyi84} analyzed the magnetoresistance that is due to electron-electron interactions for a 2D gas with smooth disorder in the ballistic regime $T\tau_{imp} \gg 1$, where $\tau_{imp}$ is the elastic scattering time ($k_B=\hbar=1$). They obtained that the magnetoresistance scales as $\omega_c^2 T^{-1/2}$ in strong fields $\omega_c \gg T$ and is exponentially suppressed at $\omega_c \ll T$, where $\omega_c$ is the cyclotron frequency. 
%
%
Very recently, Sedrakyan and Raikh \cite{Sedrakyan08} studied the ballistic regime for a 2D gas with short-range impurity potential. {They obtained that in weak fields $\omega_c E_F^{1/2}/T^{3/2} \ll 1$,  a positive magnetoresistance scales as $\omega_c^2/T^{3/2}$. In strong fields $\omega_c E_F^{1/2}/T^{3/2} \gg 1$ it crosses over to a temperature-independent negative one and scales as $\omega_c$. However these theories strongly rely on the presence of impurities in the system, while in the experiments of Renard et al., the electron mean free path was at least 50 times larger than the size of the contact that determined its resistance.}


The translational invariance is {violated} not only in disordered systems.
In a ballistic system with restricted geometry it  is broken as well and hence the electron--electron scattering may lead to nontrivial effects even in the absence of impurities or rough boundaries. Unlike the case of a conductor with impurities, these effects can be captured even in the semiclassical approximation. Very recently it was shown that in semiclassical ballistic 2D contacts with a large number of transverse channels this scattering results in a positive correction to the Sharvin conductance that scales linearly with temperature.\cite{Nagaev08}
%
The correction to the conductance results from collisions of electrons in the leads
incident on the contact with nonequilibrium electrons injected from the opposite electrode. {Bring to
notice that this mechanism gives a positive {contribution} to {the} conductance unlike the 
backscattering of electrons with few quantum channels, which results in negative corrections to {it}.\cite{Matveev04,Kindermann06,Meidan05,Syljuasen07,Rech08,Rech09,Lunde06,Sablikov06,Sloggett09,Lunde09}}
Here we analyze the behavior of this correction in a magnetic field and find that it results in an anomalous positive low-field magnetoresistance.

In this paper, we show that the positive contribution to the conductance from electron--electron scattering is effectively suppressed by a magnetic field transverse to the contact plane. This suppression results in a positive low-field magnetoresistance. After the interaction contribution has been destroyed by the magnetic field, the positive magnetoresistance gives way to a negative one that arises from the suppression of geometrical backscattering.

The paper is organized as follows. In Section II, we present the model and describe our general formalism. In Section III, we qualitatively discuss the interaction correction to the conductance in a zero magnetic field. In Section IV, we present calculations of the magnetoresistance. Section V presents a discussion the results and their comparison with the experiment. Appendices A and B contain details of calculations.

\section{Model and basic equations}

We adopt the model of a ballistic contact similar to that of Kulik {\it et al.} \cite{Kulik77} for the case of electron--phonon scattering. Consider two 2D electron gases separated by a thin impenetrable barrier with a gap of width 
$2a$. We assume that $a$ is much larger than the Fermi wavelength and the screening radius but much smaller than both elastic and inelastic mean free path of electrons. The distribution functions $f(\bp,\br)$ of electrons on both sides of the insulator obey the Boltzmann equation{\cite{Lifshitz}}
\be
 \bv\,\frac{\partial f}{\partial\br}
 +
 \left( e{\bf E}+ \frac{e}{c}\,{\bf v}\times {\bf H}\right)
 \frac{\partial f}{\partial\bp}
 =
 \hat{I}_{ee}(\bp,\br),
 \label{Boltz}
\ee
where ${\bf E} = -\nabla\varphi$ is the electric field and $\bf H$ is the magnetic field ($e<0$). The electron--electron
collision integral is given by the standard expression
\begin{eqnarray}
 \hat{I}_{ee}(\bp)
 =
 \alpha_{ee}\,\nu^{-2}
 \int\frac{d^2k}{(2\pi)^2}
 \int\frac{d^2p'}{(2\pi)^2}
 \int d^2k'\nonumber\\
 \times
 \delta(\bp + \bk - \bp' - \bk')\,
 \delta( \eps_{\bp} + \eps_{\bk} - \eps_{\bp'} - \eps_{\bk'} )
\nonumber\\
 \times
 \Bigl\{
  [1 - f(\bp)]\,[1 - f(\bk)]\, f(\bp')\, f(\bk')\,
\nonumber\\
  -
  f(\bp)\, f(\bk)\, [1 - f(\bp')]\, [1 - f(\bk')]
 \Bigr\},
 \label{I_ee}
\end{eqnarray}
where $\alpha_{ee}$ is the dimensionless parameter of electron--electron scattering\cite{screening} and $\nu= dn_s/dE_F=m/\pi$ is the two-dimensional Fermi density of states. The coordinate $\br$ is omitted in all the arguments for brevity. We assume that the electric potential $\varphi$ tends to $V/2$  and  $-V/2$ far from the contact in the left and right half-spaces, where $V$ is the voltage drop across the contact. We also assume that the distribution function of electrons in the momentum space tends to the equilibrium one sufficiently far from the contact. 

We emphasize that unlike a number of authors who considered contacts in a form of a long and narrow channel {\cite{Matveev04,Kindermann06,Meidan05,Syljuasen07,Rech08,Rech09,Lunde06,Maslov95}} with interaction effects taking place deep inside them, we consider a short and wide contact. Therefore the most essential scattering events occur outside it in the leads. 

We briefly discuss here the relative roles of the collision integral and the equilibrium boundary conditions  for the distribution function. 
In the Landauer--B\"uttiker formalism  the dissipation of power in a contact with a perfect transmission is due to relaxation processes in the leads. This relaxation brings the injected electrons in equilibrium with those in the electrodes. It is implicitly taken into account by assuming that electrons incident on the contact from the electrodes have an equilibrium distribution. However this implicit relaxation  does not take into account the back action of injected nonequilibrium electrons upon  the electrons in the leads, which may give rise to nontrivial effects. Though collisions between electrons do not change their total momentum, they change their trajectories and hence may prevent some of them from passing through the contact or help some extra electrons to get through.

In our semiclassical formalism, we also use the equilibrium boundary conditions for the electrons moving to the contact from the depth of the electrodes. If we set the collision integral in the Boltzmann equation equal to zero, we recover the Sharvin conductance, i.e. the same result as in the Landauer--B\"uttiker formalism. If the collision integral is explicitly taken into account, it not only results in a relaxation of nonequilibrium injected electrons, but also changes the distribution of ``native'' electrons coming from the depth of the electrodes. In particular, collisions with electrons injected from the other electrode make the electrons incident on the contact nonequilibrium even before they reach the orifice.  This is a sort of an electron--electron drag, which results in a correction to the Sharvin resistance. 

\begin{figure}[t]
 \includegraphics[width=8.5cm]{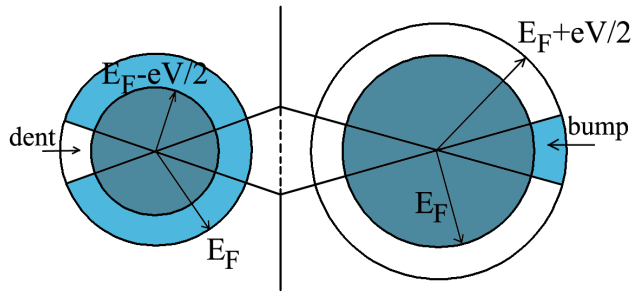}
 \caption{\label{fig1} (Color online) The distributions of non-interacting electrons to the left and to the right of the orifice  
 \cite{Kulik77}.
 In the left half-space, the Fermi sphere has a dent because the electrons injected from the contact have lower  
 energies.  In the right half-space, the Fermi sphere has a bump because the electrons injected from the contact have  
 higher  energies.
 }
\end{figure}

Equation (\ref{Boltz}) may be solved by expanding it in powers of $\alpha_{ee}$.
In the absence of scattering, electrons move along the classical trajectories in the phase space {so that their
coordinate $\br_{\tau}$ and momentum $\bp_{\tau}$ are}
determined by equations
\be
 \frac{d\bp_{\tau}}{d\tau} = e{\bf E}(\br_{\tau}) + \frac{e}{c}\,\bv_{\tau}\times{\bf H},
 \qquad
 \frac{d\br_{\tau}}{d\tau} = \frac{\bp_{\tau}}{m},
 \label{paths}
\ee
where $\tau$ is the travel time. {The total energy of an electron $p^2/(2m) + e\varphi(\br)$ is conserved during its motion along the trajectories (\ref{paths}). The boundary conditions for Eq. (\ref{Boltz}) take a form $f(\bp,\br) = f_0(\eps_{\bp})$  in the left and right half-spaces far from the orifice, where 
\be
 f_0(\eps_{\bp}) = \frac{1}{1 + \exp(\eps_{\bp}/T)}
 \label{f0}
\ee
and $\eps_{\bp}=p^2/(2m)-E_F$}.
Because of the energy conservation, the distribution function $f(\bp, \br)$ depends only on the electrode from which an electron with momentum $\bp$ came to point $\br$. As the electrons that contribute to the current belong to a narrow interval of energies of the order of $T \ll E_F$, we may treat their velocity as energy independent and  assume that the trajectories of the relevant electrons depend only on the momentum direction and do not depend  {on their energy}. It is convenient to use the notion of {an angular domain} $\Omega(\br)$ that contains all the momenta of electrons that came to point $\br$  from the contact. In terms of this {domain}, the zero-approximation distribution function 
is{\cite{Nagaev08,Kulik77,Andreev06}}
\be
 f^{(0)}(\bp, \br) =
 \left\{
  \begin{array}{ll}
   f_0({\eps_{\bp}+e\p(\br)} \mp eV/2), & \bp \notin \Omega(\br) \\
   f_0({\eps_{\bp}+e\p(\br)} \pm eV/2), & \bp \in    \Omega(\br)
  \end{array}
 \right.
 \label{f_zero}
\ee
for the electrons in left (upper sign) and right (lower sign) half-spaces, respectively. Schematically, the electron
distribution functions to the left and to the right of the contact are shown in Fig. \ref{fig1}.

The current through the contact is given by an expression
\be
 I=
 e\int d\bm{\rho}\,
 \int\frac{d^2p}{(2\pi)^2}\,
 v_{\perp}\, f(\bp,\bm{\rho}),
 \label{current}
\ee
where $v_{\perp}$ is the component of $\bf v$ normal to the insulator and vector $\bm{\rho}$ labels points within the gap in the plane of insulator. A substitution of Eqs. (\ref{f_zero}) into this expression results in the well known formula for the Sharvin conductance
\be
 G_0 = \frac{e^2 p_F a}{\pi^2},
 \label{G_0}
\ee
i.e. the conductance quantum times the number of transverse  channels in the contact.

To the first approximation in $\alpha_{ee}$, the correction to the distribution function $\delta f(\bp, \brho)$ at the orifice is readily obtained by integrating $\hat{I}_{ee}\{f^{(0)}\}$ along the classical trajectory of an electron  that arrives at point $\brho$ with momentum $\bp$ from infinity
\be
 \delta f(\bp,\brho) 
 =\int_0^{\infty} d\tau\, I_{ee}\{f^{(0)}(\bp_{\tau},\br_{\tau})\},
 \label{df}
\ee
where $\tau$ is the time of travel to point $\brho$ along the trajectory and $\bp_{\tau}$ and $\br_{\tau}$ obey Eqs. (\ref{paths}). The collision integral in Eq (\ref{df}) is zero if there is no voltage drop across the contact and the distribution function is equilibrium. Therefore as we are interested in a linear response to the electric field, we may neglect it in Eqs. (\ref{paths}) and assume that $|\bp|=$ const. In a magnetic field, the trajectories in the coordinate space present arcs of circles of cyclotron radius $l_H = p_F c/eH$, while the momentum direction $\bp_{\tau}$ rotates about the origin with an angular frequency $\omega_c = eH/mc$. {Similarly, one can neglect $\p(\br)$ in Eq.(\ref{f_zero}).}

The collision integral in Eq. (\ref{df}) is nonzero only if at least one of the momenta in Eq. (\ref{I_ee}) falls within $\Omega(\br_{\tau})$. As we will see below, the main contribution to (\ref{df}) comes from points $\br$ located much farther from the orifice than its size $a$. Hence $\Omega(\br)$ may be considered as small and the contribution to (\ref{df}) from scattering processes where more than one momentum lies in $\Omega(\br)$ may be neglected. As we will see in the next section, the largest contribution to $I_{ee}$ comes from  the collisions of electrons incident on the orifice with electrons that are injected from the other half-plane and have nearly opposite momentum. Hence the integration over $\bk$ in Eq. (\ref{I_ee}) in the left and right half-planes may be limited  to $\bk \in \Omega(\br_{\tau})$. The electrons with momentum $\bk$ should be considered as injected, and the electrons with the rest of momenta $\bp$, $\bp'$, and $\bk'$, as native to the corresponding half-plane. Therefore the collision integral, e. g., in the left electrode may be written in a form
\be
I_{ee}(\bp_{\tau},\br_{\tau})
 = \alpha_{ee}\, \nu^{-2} \int\frac{d^2k}{(2\pi)^2}
 \int\frac{d^2p'}{(2\pi)^2} \int d^2k'\,
\nonumber\\
 \times
 \delta(\bp_{\tau} + \bk - \bp' - \bk')\,
\nonumber\\
 \times
 \delta(\eps_{\bp} + \eps_{\bk} - \eps_{\bp'} - \eps_{\bk'})\,
 \Theta[\bk \in\Omega(\br_{\tau})]
\nonumber\\
 \times
 \int d\eps \int d\eps' \int d\eps_1 \int d\eps_2\,
\nonumber\\
 \times
 \delta( \eps_{\bp} - \eps )\,
 \delta( \eps_{\bk} - \eps' )\,
 \delta( \eps_{\bp'} - \eps_1 )\,
 \delta( \eps_{\bk'} - \eps_2 )\,
\nonumber\\
 \times F_L(\eps, \eps', \eps_1, \eps_2),
 \label{I_ee-2}
\ee
where
\be
 F_L(\eps, \eps', \eps_1, \eps_2)
\nonumber\\
 = [1 - f_{L}(\eps)]\,[1 - f_{R}(\eps')]\,f_{L}(\eps_1)\,f_{L}(\eps_2)
\nonumber\\
  -
  f_{L}(\eps)\,f_{R}(\eps')\,[1 - f_{L}(\eps_1)]\,[1 - f_{L}(\eps_2)].
 \label{F-def}
\ee
For convenience, we introduced here extra variables $\eps$, $\eps'$, $\eps_1$, and $\eps_2$ to separate
integrations over the energies and momentum directions {and used a notation $f_L(\eps) = f_0(\eps-eV/2)$ 
and $f_R(\eps) = f_0(\eps+eV/2)$.}
By substituting the collision integral into (\ref{df}) and the resulting expression into (\ref{current}), one obtains 
the correction to the current in a form
\be
 \delta I
 =2\alpha_{ee}\,\nu^{-2}
 \int d\eps \int d\eps' \int d\eps_1 \int d\eps_2\,
\nonumber\\
 \times
 \delta(\eps + \eps' - \eps_1 - \eps_2)\,
 F_L(\eps, \eps', \eps_1, \eps_2)
\nonumber\\
 \times
 \int\frac{d^2p}{(2\pi)^2}\, v_{\perp}\,\Theta(v_{\perp})\,
 \delta( \eps_{\bp} - \eps )\,
\nonumber\\
 \times
 \int\, d\brho \int_0^{\infty} d\tau\,
 \int\frac{d^2k}{(2\pi)^2}\,\delta( \eps_{\bk} - \eps' )\,
 \Theta[\bk \in\Omega(\br_{\tau})]\,
 \nonumber\\
 \times
 A(\eps_1, \eps_2, |\bp_{\tau} + \bk|),
 \label{dI-4}
\ee
where the quantity
\be
 A(\eps_1,\eps_2, |\bp_{\tau} + \bk|)= 
 \frac{1}{(2\pi)^2}
 \int{d^2p'}\int d^2k'\, 
\nonumber\\
 \times
 \delta(\bp' + \bk' - \bp_{\tau} - \bk)\,
 \delta(\eps_{\bp'}  - \eps_1 )\,
 \delta(\eps_{\bk'} -\eps_2 ).
 \label{A_def}
\ee
presents the effective volume of momentum space into which a pair of electrons  with momenta
$\bp$ and $\bk$ can scatter upon a collision. The prefactor 2 in Eq. (\ref{dI-4}) is due to a 
summation of two equal contributions from the left and right half-planes.

In Eq. (\ref{dI-4}), it is convenient to make the argument of $A$ independent of $\tau$. To this end,
we rotate the local coordinate system in such a way that $\bp_{\tau}$ coincides with $\bp$. As we do it,
the domain of integration over $\bk$ changes from $\Omega(\br_{\tau})$ to $\tilde\Omega(\br_{\tau},\bp_{\tau})$.
Change now the sequence of integrations in such a way that integrations over $\tau$ and $\brho$ take place first.
We arrive at an expression
\be 
 \delta I =
 4e a\, \alpha_{ee} \nu^{-2}\!\!
 \int\! d\eps\!\! \int\! d\eps'\!\! \int\! d\eps_1\!\! \int\! d\eps_2\,
\nonumber\\ 
 \times
 \delta(\eps + \eps' - \eps_1 - \eps_2)\,
 F_L(\eps,\eps',\eps_1,\eps_2)\,
\nonumber\\
 \times
 \int\frac{d^2p}{(2\pi)^2}\,
 \delta(\eps_{\bp} - \eps)\,\Theta(v_{\perp})\,v_{\perp}
\nonumber\\ 
 \times
 \int\frac{d^2 k}{(2\pi)^2}\,
 \delta(\eps_{\bk}-\eps')\, 
 A(\eps_1,\eps_2, \bp + \bk)\,
 \bar{\tau}_m(\bp,\bk),
 \label{dI-5}
\ee
where 
\be 
 \bar\tau_m(\bp,\bk) =
 \frac{1}{2a}\int d\rho \int\limits_0^{\infty} d\tau\,
 \Theta[\bk \in\tilde\Omega(\br_{\tau},\bp_{\tau})]
 \label{tau_def}
\ee
is the effective time of interaction between incident electrons with momentum $\bp$ and injected 
electrons with momentum $\bk$.

\section{Zero magnetic field: a qualitative discussion}

\begin{figure}[t]
 \includegraphics[width=8.5cm]{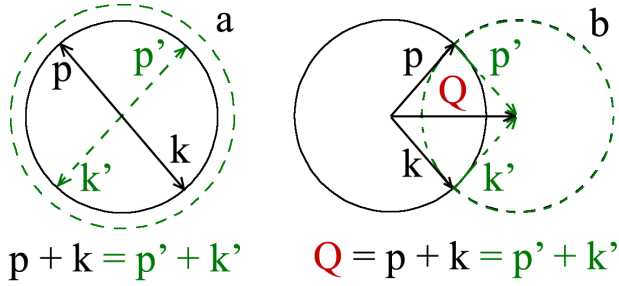}
 \caption{\label{fig2} (Color online) Restrictions in the momentum space for electron scattering. (a) Electrons with oppositely
 directed velocities have an infinite number of possibilities to
 scatter in a pair of electrons with oppositely directed velocities and give the resonant contribution to the
 conductivity. (b) Electrons with nonzero total momentum $Q$ may only exchange either retain their momenta.
 }
\end{figure}

First consider qualitatively the case of zero magnetic field. In the limit of zero magnetic and electric 
field, the electrons just move along straight lines with a constant momentum. The interaction correction to the 
conductance for this case was calculated in Ref. \onlinecite{Nagaev08}. The expression for this correction had 
a form
\be
 \frac{\delta G}{G_{0}} =
 \frac{2C_0}{\pi^2}\,\alpha_{ee}\,
 \frac{a}{v_F}\, T\,
 \ln\left(\frac{l_c}{a}\right),
 \label{dG_2D}
\ee
where $C_0 \approx 1.87$ and $l_c$ is a cutoff length much larger than $2a$, which may be due to a very weak electron--impurity
scattering or a finite size of the electrodes. In contrast to the case of a long and narrow 
contact,{\cite{Matveev04,Kindermann06,Meidan05,Syljuasen07,Rech08,Rech09,Lunde06,Sablikov06}} 
the correction is positive. The unusual sign of the correction may be explained as follows. In the absence of 
interaction, the electron distribution in the momentum space has a dent or a bump 
at the side opposite to the contact 
depending
on the electrode (see Fig.\ref{fig1}). The electron--electron scattering tends to smooth out this dent or bump by adding electrons to that part of the Fermi surface or removing them from there. As the total momentum of the electron
gas must be conserved, the center of mass of the local distribution should not be changed by the scattering. Hence
additional electrons should also appear at the side facing the contact or be removed from there. As  a result, the number of incident electrons increases  at the negative-voltage side of the contact and decreases at the positive-voltage side of it thus forming a positive correction to the current. 

\begin{figure}[t]
 \includegraphics[width=8.5cm]{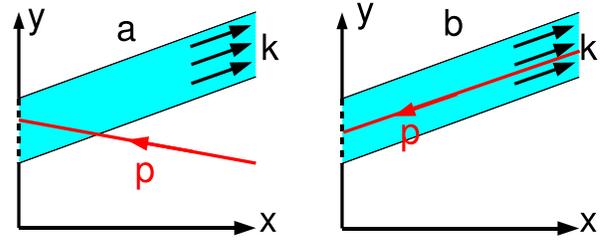}
 \caption{\label{fig3} (Color online) Effective time of electron scattering in zero magnetic field. (a) If 
 the momentum direction
 of an incident electron differs from that of injected electrons, the time of their interaction is limited. (b) If the
 momentum of an incident electron is opposite to that of injected electrons, the time of their interaction is
 infinite.
 }
\end{figure}

The magnitude of the interaction correction is much larger than one would expect from qualitative considerations. Naively, one 
might expect $\delta G/G_0$ to be of the order of $a/l_{ee}$, where $l_{ee}^{-1} \sim T^2/(v_F E_F)$ is the inverse equilibrium electron--electron scattering length. However the actual effect is $(E_F/T)\ln(l_c/a)$ times larger. The large magnitude of the effect is due to a special role played by collisions of electrons with almost opposite momenta.\cite{Gurzhi95} A sharp increase in the volume of momentum space available for scattering upon such collisions results in a divergence of 
quantity $A$ in Eq. (\ref{dI-5}) at $\bp + \bk=0$. A qualitative explanation of this singularity is given in Fig. \ref{fig2}. Imagine that the two initial states of scattering electrons $\bp$ and $\bk$ lie exactly at the Fermi surface and the temperature is zero. Because of energy conservation and Fermi statistics, the final scattering states $\bp'$ and $\bk'$ should also lie exactly at the Fermi surface and obey the momentum conservation condition $\bp + \bk = \bp' + \bk'$. If ${\bf Q} = \bp + \bk \ne 0$, this condition may be satisfied only if the electrons retain their momenta or exchange them. However if ${\bf Q}=0$, the electrons can scatter into an arbitrary pair of states with opposite momenta and hence the amount of available volume in the momentum space drastically increases.

If the electrons have a smooth angular distribution, the singularity in $A(|\bp+\bk|)$ is averaged over the whole Fermi surface and eventually results only in an additional logarithmic factor $\ln(E_F/T)$ in the electron--electron scattering rate.\cite{Hodges71,Giuliani82} However in our problem, $A(|\bp+\bk|)$ is averaged over the Fermi surface with a weight factor $\bar\tau_m(\bp,\bk)$, which accounts for the anisotropy of local electron distribution and presents the dwell time of an electron with momentum $\bp$ incident on the contact in a beam of electrons injected with momentum $\bk$ (see Fig. \ref{fig3}). This time remains finite if $\bp$ is tilted with respect to $\bk$, but it tends to infinity if $\bp=-\bk$. In the latter case, the incident and injected electrons move along the same straight line in the opposite directions and one can find a nonequilibrium electron with momentum $-\bp$ no matter how far from the orifice. The superposition of the two singularities in $A(|\bp+\bk|)$ and $\bar\tau_m(\bp,\bk)$ in the integrand of Eq. (\ref{dI-5}) results in an extra power of $E_F/T$ in the final expression for the current (\ref{dG_2D}).

\begin{figure}[t]
 \includegraphics[width=8.5cm]{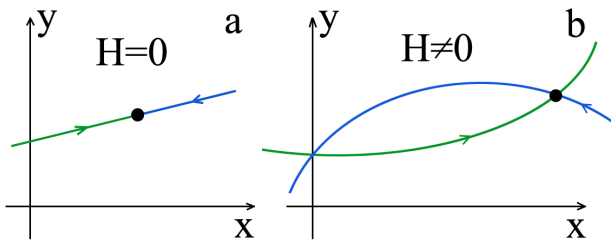}
 \caption{\label{fig4} (Color online) Destruction of the resonance of oppositely moving electrons in magnetic field. 
(a) Electrons move oppositely 
and give a resonant contribution. (b) Electron trajectories are bent by the
magnetic field and  the electron velocities
form a nonzero angle at the collision point.
 }
\end{figure}

The ``resonance'' scattering  of electrons with opposite momenta is due to the time-reversal symmetry, which makes them move along the same trajectory in opposite directions. If this symmetry is destroyed by a magnetic field, the effect is strongly suppressed. The magnetic field bends trajectories of electrons with opposite momenta in opposite directions so that they do not coincide any more (see Fig \ref{fig4}). This leads to a reduction of the effect of scattering on the conductance and hence to a positive magnetoresistance. In the next sections, we give a quantitative estimate of the related magnetoresistance.

\section{Calculation for nonzero magnetic field}

\begin{figure}[t]
 \includegraphics[width=8.5cm]{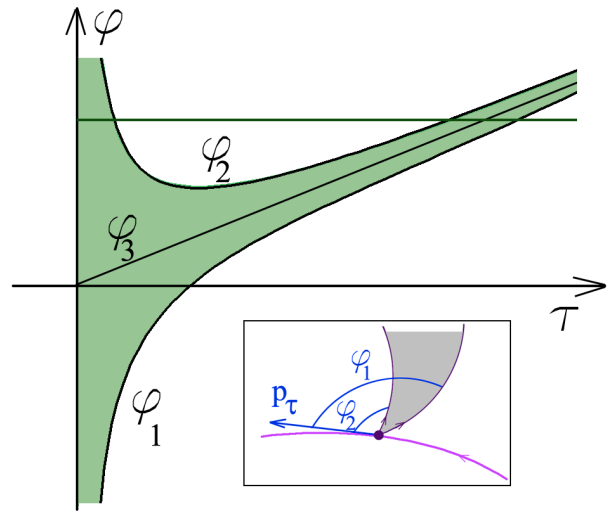}
 \caption{\label{fig5} (Color online) Dependencies of the angles $\p_1$ and $\p_2$ between the momenta of an incident 
 electron and electrons injected from the upper and lower edges of the gap
 on the travel time of incident electron to the contact $\tau$. $\p_3$ presents the same dependence for
 an electron injected at $y'=y$. {The time $\tau_m(\p)$ presents the sum of the portions of the line 
 $\p={\rm const}$ lying within the shaded area. The inset shows the mutual arrangement of the trajectories
 of the incident and the injected electrons at the collision point $\br_{\tau}$.
 The shaded area shows $\Omega(\br_{\tau})$ and contains the momentum directions of injected electrons.}
 }
\end{figure}

The anomalously large interaction correction to the current results from collisions that take place at large distances from the contact $v_F\tau \gg a$. Therefore the divergence in $\bar\tau_m(\bp,\bk)$ should be suppressed already in 
weak magnetic fields such that $l_H \gg a$. This is why in what follows we will be interested in low magnetic fields and large $\tau$.

\begin{figure}[t]
 \includegraphics[width=8.5cm]{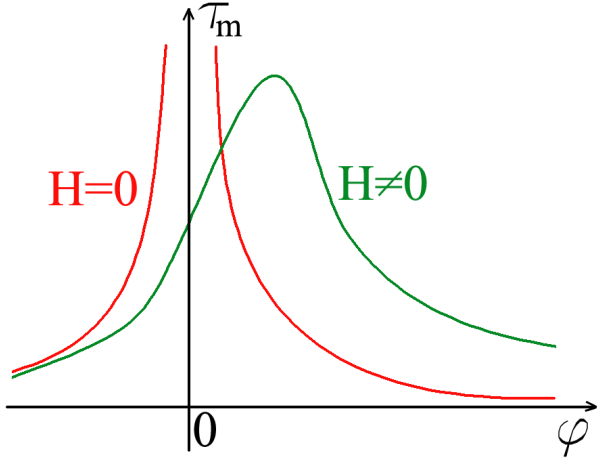}
 \caption{\label{fig6} (Color online) Dependence $\tau_m(\p)$ for the cases of the zero and nonzero magnetic field.
 }
\end{figure}

To calculate the effective time $\bar\tau_m$, one should first obtain the angular boundaries of the  
domain of integration $\Omega( \textbf{r}_\tau, \textbf{p}_\tau)$ for arbitrary $\tau$. For 
this purpose, one has to solve Eqs. (\ref{paths}) in the limit of zero electric field. 
Introduce a coordinate system with the origin at the center of the contact and the $y$ axis parallel to the insulating layer. 
Consider a specific trajectory of an electron incident on the contact. Suppose that the trajectory crosses the contact at point $\brho=(0,y)$ at an angle $\varphi_p$ to the contact normal. Determine the boundary angles for integration over $\bk$ at any point of the trajectory $\br_{\tau}$ corresponding to travel time to the contact $\tau$.  If another electron is injected from the contact at point $\brho'=(0,y')$ and then arrives at point $\br_{\tau}$, its momentum will make 
an angle $\varphi$ with the momentum $\bp_{\tau}$ that the incident electron had at this point. In the linear approximation in the magnetic field and at $\tau \gg a/v_F$ this angle is given by an asymptotic expression (see Appendix A)
\be
 \p(\tau)=\omega_c\tau + \frac{y-y'}{v_F\tau} \cos\p_p.
 \label{phi}
\ee
In the $(\tau,\p)$ plane, the region $\bk \in\tilde\Omega(\br_{\tau},\bp_{\tau})$  is limited by two curves $\p_1(\tau)$ and $\p_2(\tau)$, which are obtained by substituting $y'=a$ and $y'=-a$ into (\ref{phi}) and correspond to electrons injected from the upper and lower edges of the gap (see Fig. \ref{fig5}). The effective interaction time $\bar\tau_m(\p,\p_{\bp})$ presents the total length of the portions of the line $\p$ = const that fall inside this region. This length has to be integrated over $y$ from $-a$ to $a$ and divided by $2a$. Calculations give that (see Appendix B)
\be
 \bar\tau_m(\p,\p_p)
 =\frac{a}{v_F}\,\,\sqrt{\frac{\cos\p_p}{\beta}}\,
 G\!\left(\frac{\p}{\sqrt{4\beta\cos\p_p}}\right),
 \label{tau_m-expl}
\ee
where $\beta = \omega_c a/v_F$ is dimensionless magnetic field and
\be
 G(b)=
 \begin{cases}\Dsp
  b + \frac{1}{3}\,(b^2 + 2)^{3/2} + \frac{1}{3}\,b^3, & b<0           \\ \Dsp
  b + \frac{1}{3}\,(b^2 + 2)^{3/2} - b^3,              & 0<b<\sqrt{2}  \\  \Dsp
  b + \frac{1}{3}\,(b^2 + 2)^{3/2}                     & \quad         \\ \dsp
  + \frac{2}{3}\,(b^2 - 2)^{3/2}- b^3, & b>\sqrt{2}.
 \end{cases}
\label{G}
\ee
The obtained piecewise function is shown by the green continuous line in Fig. \ref{fig6}.
The magnetic field not only smooths out the singularity at $\p=0$ that corresponds to the case of oppositely moving electrons but also shifts the maximum of $\bar\tau_m(\p)$ away from this point. The $\bar\tau_m(\p)$ dependence  for the case of zero magnetic field is shown in this figure by red lines. 

Quantity $A$ that also enters Eq. (\ref{dI-5}) presents an integral over the Fermi surface over two momenta with a given sum and reflects the limitations on the scattering in the phase space that result from momentum conservation. An explicit calculation gives
\be
 A(\eps_1,\eps_2, \bp + \bk) 
 = \frac{1}{v_F^2} \frac{1}{(2\pi)^2}
\nonumber\\
 \times
 \frac
 {\Theta\!\left(\sin^2\frac{\varphi}{2} + \frac{D}{4E_F^2}\right)}
 {\cos(\varphi/2)\sqrt{\sin^2\frac{\varphi}{2} + \frac{D}{4E_F^2}}},
 \label{A-expl}
\ee
where
\be
 D = \frac{(\eps_p - \eps_k)^2 - (\eps_2-\eps_1)^2}{4}.
 \label{D}
\ee
Upon integration over $\bp$ and $\bk$ in Eq. (\ref{dI-5}) it reduces to an integral over energies
\be
 \delta I = 4e a \frac{\alpha_{ee}}{\nu^2}\int{d\eps\int{d\eps'\int{d\eps_1\int{d\eps_2}}}} 
\nonumber\\
 \times 
 \delta(\eps + \eps' - \eps_1 - \eps_2)\,F_L(\eps, \eps', \eps_1, \eps_2)\,
 C(\eps, \eps', \eps_1, \eps_2),
 \label{dI-eps}
\ee
where $C(\eps, \eps', \eps_1, \eps_2)$ can be found explicitly in several cases. If the ratio
$D/(4E_F^2\beta)$ is large and positive,
\be
 C = C_1 = \frac{\nu^2}{8\pi^3}\,
 \frac{a}{v_F^2}\,
 \frac{E_F}{\sqrt{D}}\,
 \ln\!\left(\frac{D}{4\beta E_F^2}\right).
 \label{>>1}
\ee
If this ratio is large and negative, 
\be
 C = C_2 = \frac{\nu^2}{8\pi^2}\,
 \frac{a}{v_F^2}\,
 \frac{E_F}{\sqrt{\mid D \mid}}.
 \label{<<-1}
\ee 
If the absolute value of this ratio is small,
\be
 C = C_3 = 
 -\frac{\nu^2}{3\pi^4}\,
 \frac{a}{v_F^2}\,\frac{1}{\sqrt{\beta}}\,
 \ln\!\left|\frac{D}{4\beta E_F^2}\right|.
 \label{<<1}
\ee

The distribution-dependent factor $F_L$ in (\ref{dI-eps}) should be linearized with respect to the voltage.
With account taken of the delta function in this equation it may be written in a form
\be
 F_L(\eps,\eps',\eps_1,\eps_2) =
 \frac{eV}{T}\,
 \exp\!\left({\frac{\eps+\eps'}{T}}\right)\,
\nonumber\\
 \times
 f_0(\eps)\,f_0(\eps')\,f_0(\eps_1)\,f_0(\eps_2).
 \label{F-lin}
\ee
This factor exponentially falls off away from the Fermi surface and limits the integration over the energies 
to a narrow interval of width $T$ near it. Hence one may use an estimate
$$
 \frac{D}{4E_F^2\beta} \sim \frac{T^2}{4E_F^2\beta}.
$$
Equation (\ref{dI-eps}) is easily evaluated  in two limiting cases of weak and strong magnetic fields by introducing dimensionless variables $\xi_i = \eps_i/T$.  In the case of a weak magnetic field $\beta \ll T^2/E_F^2$ the most singular in $T$ contribution to the integral is 
given by negative values of $D$, and a substitution of (\ref{<<-1}) into (\ref{dI-eps}) gives
\be
 \frac{\delta G}{G_{0}} = \frac{C_0}{8\pi}\alpha_{ee}\frac{a}{v_F}T
 \ln\!\left(\frac{T^2}{\beta\,E_F^2}\right).
 \label{low}
\ee
This equation has the same form as (\ref{dG_2D}) with $l_c = (T/E_F)^2\,l_H$ except for the numerical prefactor.
The different numerical prefactors in Eqs. (\ref{dG_2D}) and (\ref{low}) show that a true crossover between the cases 
of $H=0$ and $H \ne 0$ takes place only at $l_H \gg l_c$.

In the opposite case of a relatively strong magnetic field $T^2/E_F^2 \ll \beta \ll 1$ a substitution of (\ref{<<1})
into (\ref{dI-eps}) gives with a logarithmic accuracy
\be
 \frac{\delta G}{G_{0}} =
 \frac{4}{9}\alpha_{ee}\frac{a}{v_F}
 \frac{T^2}{E_F \sqrt{\beta}}
 \ln\!\left(\frac{\beta E_F^2}{T^2}\right).
 \label{high}
\ee
This equations suggests that the correction to the conductance is positive, increases with temperature 
and decreases with magnetic field. At higher magnetic fields, the interaction correction to the conductance 
falls off more rapidly.

\section{Discussion}

\begin{figure}[t]
 \includegraphics[width=8.5cm]{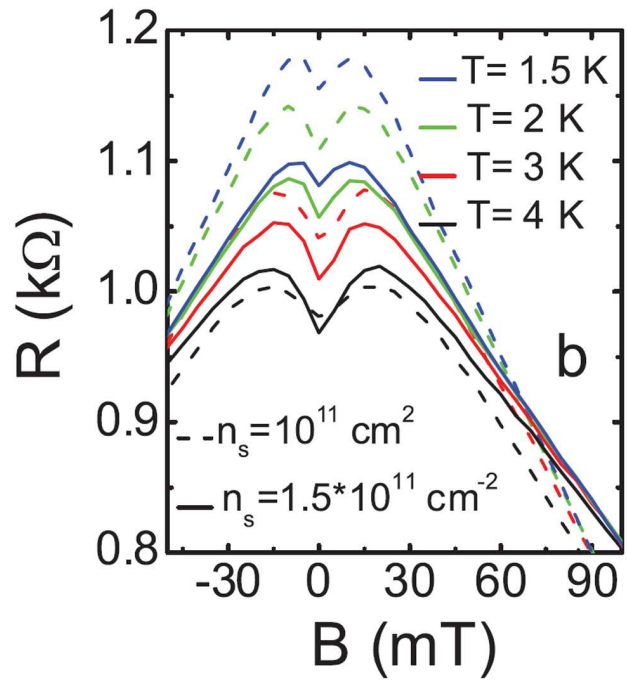}
 \caption{\label{fig7} (Color online) Experimental four-terminal magnetoresistance for a ballistic 2D contact with 13 open 
 quantum channels.\cite{Renard08}
 }
\end{figure}

A positive correction to the conductance that decreases with increasing magnetic field results in a positive low-field magnetoresistance $\delta R(H) = -\delta G/G_0^2$.
The low-temperature portions of these curves are similar to those of experimental curves obtained by Renard et al.\cite{Renard08} (see Fig. \ref{fig7}). In both cases, higher temperatures correspond to lower $R(H)$.

At higher magnetic fields the experimental magnetoresistance exhibits a maximum and eventually decreases linearly with $H$. This linear decrease may be attributed 
to the geometrical suppression of electron backscattering from the contact\cite{Beenakker91} in a four-terminal geometry. The four-terminal resistance for noninteracting electrons is given by a formula\cite{vanHouten88}
\be
 R_4(H) = R_2 - \frac{2\pi}{e^2}\,\frac{1}{p_Fl_H},
 \label{4t}
\ee

\noindent
where $R_2$ is independent of $H$ at $l_H \gg a$. For interacting electrons, the maxima in the curves can be explained by a crossover from the positive magnetoresistance governed by interaction effects to the negative magnetoresistance related with the suppression of geometrical backscattering. One may roughly estimate the positions of maxima in 
$R_4(H)$ by substituting $R_2 = G_0^{-1} - \delta G/G_0^2$ into Eq. (\ref{4t}) and differentiating it with respect to $H$. Estimates made for a 2DEG on the basis of a GaAs heterostructure with electron concentration $n_s = 1.5\times10^{11}$ cm$^{-2}$, $T=1.5$ K, $\alpha_{ee}=1$,\cite{estimate} and a contact with 13 open quantum channels\cite{Renard08} give the maximum of
$R_4(H)$ at a magnetic field about 10 mT. This is 
in a good agreement with the experimental data (see Fig. \ref{fig7}). The behavior of the experimental curves also qualitatively agrees with our results. As the temperature increases and the interaction effects become larger, the maxima become more pronounced and shift towards higher fields. At sufficiently high fields where the interaction correction is completely suppressed, all the curves merge into a single temperature-independent straight line in accordance with Eq. (\ref{4t}).

The authors of Ref. \onlinecite{Renard08} attributed the linearly dependent on temperature contribution to the zero-field conductance of the contact to an electron scattering off the Friedel oscillations around it. Our semiclassical model
also predicts an interaction correction to the conductance that linearly depends on the temperature in a zero field.\cite{Nagaev08} 
However in our opinion, the qualitative agreement 
between our calculated magnetoresistance and the experimental data suggests that the  observed features are likely to result from semiclassical electron--electron scattering. 

Though collisions of electrons with opposite momenta modify
the electron lifetime and thermodynamic properties of a homogeneous 2DEG,\cite{Chubukov03} the related anomalies 
are averaged over the whole Fermi surface and therefore these effects are difficult
to observe. Ballistic contacts in a 2DEG under voltage bias serve as selective amplifiers of the contribution
from such collisions 
and visualize them as peculiarities of the conductance.

{In 3D systems, the effects of electron--electron scattering on the conductance of ballistic contacts
are smaller by a factor $T/E_F$. Therefore the temperature-dependent magnetoresistance in them should be much 
{less pronounced} than in 2DEG.  }

In summary, we have proposed a semiclassical mechanism of magnetoresistance of ballistic contacts, which is related with a destruction of ``resonance'' scattering of oppositely moving electrons by the magnetic field. This mechanism may account for the experimentally observed peculiarities of transport.

\begin{acknowledgments}
We are grateful to M. Lunde, M. B\"uttiker, V. T. Renard, H. U. Baranger, and J.-C. Portal for useful discussions and to V. T. Renard for familiarizing us with his unpublished data. {This work was supported by Russian Foundation for
Basic Research, grants 09-02-12192-ofi-m and 10-02-00814-a, and by the  ``Quantum nanostructures'' program of Russian Academy of Sciences.}
\end{acknowledgments}

\appendix
\section {Calculation of  $\varphi(\tau)$}

Suppose that an electron incident on the contact crosses it at point $(0, y)$ a time $\tau$ after it has collided with an electron injected from the contact point $(0, y^\prime)$. Denote the collision point $\bf{r}_\tau$. 
Denote the polar angles of the momenta of the injected and the incident electrons at point $\bf{r}_\tau$ by $\varphi_k$ and 
$\varphi_{p\tau}$, so that the angle between them equals $\varphi = \varphi_k - \varphi_{p\tau}$.
The directions of the electron momenta rotate in the magnetic field, so that
\be
\varphi = \varphi_k - \varphi_p + \omega_c\tau,
\label{phi-a}
\ee
where $\varphi_p$ is the polar angle of the momentum of the incident electron at the contact.

To obtain the  $\varphi_k(\tau)$ dependence, we obtain the coordinates of collision point $\bf{r}_\tau$ in two ways by solving the equation of motion 
\be
\ddot{\bf {r}}= \frac{e}{mc}{\bf v}\times {\bf H}
\label{ext1}
\ee
for the incident and  injected electrons.
We solve this equation with boundary conditions  
\be
{\bf v}=\frac{p}{m}\left(
\begin{aligned}
\cos\p_p\\
\sin\p_p\\
\end{aligned}
\right), \;\;\;
{\bf r}=\left(
\begin{aligned}
0\\
y\\
\end{aligned}
\right)
\ee
at the gap for the incident electron and obtain the coordinates of the collision point 
\be
{\bf r}_\tau=\left(
\begin{aligned}
\frac{p}{m\omega_c}[\sin\p_p + \sin(\omega_c\tau - \p_p)]\\
y+\frac{p}{m\omega_c}[-\cos\p_p + \cos(\omega_c\tau - \p_p)]\\
\end{aligned}
\right)
\label{first}
\ee
in a first way.

For the injected electron, the outer product in Eq. (\ref{ext1}) has the opposite direction, so we should write a minus-sign before it. The boundary conditions for the injected electron at the gap are
\be
{\bf v}^\prime=\frac{k}{m}\left(
\begin{aligned}
\cos(\varphi_k - \omega_c\tau^\prime)\\
\sin(\varphi_k - \omega_c\tau^\prime)\\
\end{aligned}
\right), \;\;\;
{\bf r}^\prime=\left(
\begin{aligned}
0\\
y^\prime\\
\end{aligned}
\right).
\ee
Here $\tau^\prime$ is the time of  motion of the injected electron from the contact to $\bf{r}_\tau$.

We solve the equation of motion for the injected electron with these boundary conditions and obtain the coordinates of collision point in a second way
\begin{equation}
{\bf r}_\tau=\left(
\begin{aligned}
\frac{k}{m\omega_c}[-\sin(\p_k-\omega_c\tau') + \sin\p_k]\\
y'+\frac{k}{m\omega_c}[-\cos(\p_k-\omega_c\tau') + \cos\p_k]\\
\end{aligned}
\right).
\label{second}
\end{equation}
Equate the right-hand sides of expressions (\ref{first}) and (\ref{second}) and in approximation $p \approx k \approx p_F$, obtain a system in two unknowns $\varphi_k$ and $\tau^\prime$. We solve this system for $\varphi_k$ and then substitute it into (\ref{phi-a}) to obtain $\varphi$. Then we expand the obtained $\varphi$ into series in the small dimensionless magnetic field $\beta = a\omega_c/v_F$ and take its asymptotics at large dimensionless time of motion from the collision point to the contact $t = \tau v_F/a$, which results in an expression

\be
\p_k=t\beta + \frac{y-y^\prime}{at}\cos\varphi_p = \omega_c\tau + \frac{y-y^\prime}{v_F\tau}\cos\varphi_p.
\ee

\section{Calculation of the effective time of interaction}

\begin{figure}[t]
\includegraphics[width=8.5cm]{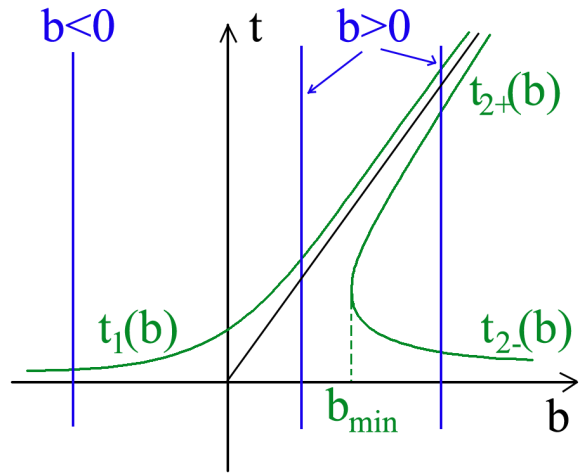}
\caption{ (Color online) Limits of integration over $t$ for different $b$.}
\label{pic_tb2}
\end{figure}

To calculate the effective time of interaction (\ref{tau_def}), one has first to determine the boundaries of the region
$\bk\in\tilde\Omega(\br_{\tau},\bp_{\tau})$ in the $(\p,\tau)$ plane. According to Eq. (\ref{phi}), they are given by
an expression
\be
 \p_{1,2}(\tau)=\omega_c\tau + \frac{y\mp a}{v_F\tau} \cos\p_p.
 \label{bound}
\ee
To perform the integration over $\tau$ in (\ref{tau_def}) and to obtain the integration limits in this equation, one has to invert these dependencies. To this end, it is convenient to introduce dimensionless variables $b = \varphi/\sqrt{4\beta\cos\p_{\bp}}$,  $t = \tau v_F/a$, and $\eta = y/a$. For electrons injected through the upper edge of the gap, the $b_1(t)$ dependence is monotone, so the inverse dependence is a single-valued function
\be
t_1(b) = \sqrt{\frac{\cos\p_p}{\beta}}\left(b + \sqrt{b^2 - (\eta - 1)}\right).
\ee
For electrons injected through the lower edge of the gap, the nonmonotone $b_2(t)$ dependence is inverted into two functions
\be
t_{2+}(b) = \sqrt{\frac{\cos\p_p}{\beta}}\left(b + \sqrt{b^2 - (\eta + 1)}\right), \\
t_{2-}(b) = \sqrt{\frac{\cos\p_p}{\beta}}\left(b - \sqrt{b^2 - (\eta + 1)}\right).
\ee

As we see in Fig. \ref{pic_tb2}, the segment borders (i.e. integration limits) are defined by different functions for different values of $b$. So there are three possible cases, which are shown in Fig. \ref{pic_tb2} by blue lines.

1. In the most simple case $b < 0$, the integration in (\ref{tau_def}) is performed over the intervals $\eta \in [-1 {,} 1]$ and  $t \in [0, t_1(b)]$, so
\be
 \bar\tau_m = \frac{a}{2v_F}
 \int\limits_{-1}^{1} d\eta
 \int\limits_{0}^{t_1(b)} dt.
 \label{c1}
\ee

2. If $b \in [\sqrt{2}, \infty)$, there are two possibilities depending on whether 
$b$ is above or below the minimum of dependence $b_2(t)$, which is located at point $b_{min} = \sqrt{\eta + 1}$. Since $\eta \le 1$, $b$ is below the minimum while in the range $b \ge \sqrt{2}$. Hence $\bar\tau_m$ assumes the form
\be
\bar\tau_m = \frac{a}{2v_F}
\left[
  \int\limits_{-1}^{b^2-1} d\eta
  \left(\int\limits_{0}^{t_{2-}(b)}{dt} + \int\limits_{t_{2+}(b)}^{t_{1}(b)}{dt}\right)  
 \right. 
\nonumber\\
  +\left. \int\limits_{b^2-1}^{1} d\eta\int\limits_{0}^{t_1(b)}{dt}\right].
 \label{c2}
\ee

3. If $b$ is in the range $b \in [0, \sqrt{2}]$, it is above the minimum in $b_2(t)$ if $\eta \in [-1 , b^2 - 1]$ and is below the minimum if $\eta \in [b^2 - 1, 1]$. So $\bar\tau_m$ is given by
\be
\bar\tau_m = \frac{a}{2v_F}\int\limits_{-1}^{1}{d\eta\left(\int\limits_{0}^{t_{2-}(b)}{dt} + \int\limits_{t_{2+}(b)}^{t_{1}(b)}{dt}\right)}.
\label{c3}
\ee

The integrations in Eqs. (\ref{c1}) - ({\ref{c3})  result in  a piecewise function (\ref{tau_m-expl}).


\begin{thebibliography}{99}
%
\bibitem{Sharvin}
Y. V. Sharvin, Zh. Eksp. Teor. Fiz. {\bf 48}, 984 (1965) [Sov. Phys. JETP {\bf 21}, 655 (1965)].
%
\bibitem{Imry}
Y. Imry, {\it Introduction to Mesoscopic Physics}, (Oxford University, New York, 2002).
%
\bibitem{Beenakker91} C. W. J. Beenakker and H. van Houten, Solid State Physics, {\bf 44}, 1 (1991).
%
\bibitem{Renard08} V. T. Renard, O. A. Tkachenko, V. A. Tkachenko, T. Ota, N. Kumada, J.-C. Portal, and Y. Hirayama, Phys. Rev. Lett. {\bf 100}, 186801 (2008).
%
\bibitem{Altshuler85} B. L. Altshuler and A. G. Aronov, in {\it
Electron-electron Interactions in Disordered Systems,} edited by A.
L. Efros and M. Pollak (North-Holland, Amsterdam, 1985), p. 1.
%
%
\bibitem{Gornyi83} I. V. Gornyi and A. D. Mirlin, Phys. Rev. Lett. {\bf 90}, 076801 (2003).
%
\bibitem{Gornyi84} I. V. Gornyi and A. D. Mirlin, Phys. Rev. B {\bf 69}, 045313 (2004).
%
\bibitem{Sedrakyan08} T. A. Sedrakyan and M. E. Raikh, Phys. Rev. Lett. {\bf 100}, 106806 (2008).
%
\bibitem{Nagaev08} K. E. Nagaev and O. S. Ayvazyan, Phys. Rev. Lett. {\bf 101}, 216807 (2008).
%
\bibitem{Matveev04}
{K.A. Matveev, Phys. Rev. Lett. {\bf 92}, 106801 (2004).}
%
\bibitem{Kindermann06}
{M. Kindermann and P.W. Brouwer, Phys.Rev. B 74, 125309 (2006).}
%
\bibitem{Meidan05}
{D. Meidan and Y. Oreg, Phys. Rev. B {\bf 72}, 121312(R) (2005).}
%
\bibitem{Syljuasen07}
{O.F. Syljuasen, Phys. Rev. Lett. {\bf 98}, 166401 (2007).}
%
\bibitem{Rech08}
J. Rech and K. A. Matveev, Phys. Rev. Lett. {\bf 100}, 066407 (2008);
{J. Phys.: Condens. Matter {\bf 20}, 164211 (2008).}
%
\bibitem{Rech09}
{J. Rech, T. Micklitz, and K. A. Matveev, Phys. Rev. Lett. {\bf 102}, 116402 (2009).}
%
\bibitem{Lunde06}
A. M. Lunde, K. Flensberg, and L. I. Glazman, Phys. Rev. Lett. {\bf 97}, 256802 (2006);
{Phys. Rev. B {\bf 75}, 245418 (2007).}
%
\bibitem{Sablikov06}
{V. A. Sablikov, JETP Lett. {\bf 84}, 404 (2006)}
%
\bibitem{Sloggett09}
C. Sloggett, A. I. Milstein and O. P. Sushkov. Eur. Phys. J. B {\bf 61}, 427 (2008)
%
\bibitem{Lunde09}
A. M. Lunde, A. De Martino, A. Schulz, R. Egger, and K. Flensberg, New J. Phys. {\bf 11},
023031 (2009)
%
\bibitem{Kulik77}
I. O. Kulik, R. I. Shekhter, and A. N. Omelyanchuk, Solid State Commun. {\bf 23}, 301 (1977).
%
\bibitem{Lifshitz}
{E. M. Lifshitz, L. P. Pitaevskii, {\it Physical kinetics, Course of theoretical physics}, 
Oxford, Pergamon Press, 1981}
%
\bibitem{screening}
We may take the interaction parameter in the collision integral as momentum-independent because of the 
short screening length. In the case of the bare Coulomb interaction this parameter would exhibit a 
singularity at small momentum transfers, and our approach would not work. 

\bibitem{Maslov95}
D. L. Maslov and M. Stone, Phys. Rev. B {\bf 52}, R5539 (1995).
%
\bibitem{Andreev06}
{A. V. Andreev and L. I. Glazman, Phys. Rev. Lett. 97, 266806 (2006)}
%
\bibitem{Gurzhi95}
The effect of such collisions in narrow channels with boundary scattering was considered by R. N. Gurzhi, A. N. Kalinenko, and A. I. Kopeliovich, Phys. Rev. Lett. {\bf 74}, 3872 (1995).
%
\bibitem{Hodges71} C. Hodges, H. Smith, and J. W. Wilkins, Phys. Rev. B {\bf 4}, 302 (1971).
%
\bibitem{Giuliani82}
G. F. Giuliani and J. J. Quinn, Phys. Rev. B {\bf 26}, 4421 (1982).
%
\bibitem{vanHouten88} H. van Houten, C. W. J. Beenakker, P. H. M. van Loosdrecht, T. J. Thornton, H. Ahmed, 
M. Pepper,  C. T. Foxon, and J. J. Harris, Phys. Rev. B {\bf 37}, 8534 (1988).
%
\bibitem{estimate}
It is very difficult to obtain a reliable estimate of the interaction constant in GaAs structures at low electron concentrations where the gas parameter is of the order of unity. If we assume that the scattering is due to the statically screened Coulomb interaction and make use of Eq. (3.29) of Ref. \onlinecite{Altshuler85}, we get $\alpha_{ee} \sim 1$.
%
\bibitem{Chubukov03}
A. V. Chubukov and D. L. Maslov, Phys. Rev. B {\bf 68}, 155113 (2003);  {\it ibid.} {\bf 69}, 121102 (2004)


\end{thebibliography}
\end{document}